# Deep Learning based Differential Distinguisher for Lightweight Block Ciphers

Aayush Jain[1], Varun Kohli[2] and Girish Mishra[3]

[1] National Institute of Technology, Tiruchirappalli, India-620015
[2] Birla Institute of Technology and Science Pilani, India-333031
[3] Scientific Analysis Group DRDO, India-110054

**Abstract.** Recent years have seen an increasing involvement of Deep Learning in the cryptanalysis of various ciphers. The present study is inspired by past works on differential distinguishers, to develop a Deep Neural Network-based differential distinguisher for round reduced lightweight block ciphers PRESENT and Simeck. We make improvements in the state-of-the-art approach and extend its use to the two structurally different block ciphers, PRESENT-80 and Simeck64/128. The obtained results suggest the universality of our cryptanalysis method. The proposed method can distinguish random data from the cipher data obtained until 6 rounds of PRESENT and 7 rounds of Simeck encryption with high accuracy. In addition to this, we explore a new approach to select good input differentials, which to the best of our knowledge has not been explored in the past. We also provide a minimum-security requirement for the discussed ciphers against our differential attack.

## 1 Introduction

In the era of Internet of Things (IoT), security is one of the deciding factors for the viability of IoT systems [1]. These include smart homes [2], medical and healthcare services [3,4], environment monitoring [5], transportation [6], vehicular networks [7], and UAVs [8] to name a few. An IoT system comprises of four layers, namely application, middleware, network and sensing [1], each of which may show vulnerability to a wide range of security threats [9]. Even seemingly secure systems have inherent flaws that can be exploited by attackers without the need of high computational resources [10]. One such flaw for example, may be biased encryption algorithms.

Various methods have been developed to exploit such flaws. Differential Cryptanalysis is a general cryptanalysis technique primarily used on block ciphers [11], with some applications in stream ciphers [12] and hash functions [13]. It is the study of how input differences affect the differences in the output. In the case of block ciphers, the technique follows the transformation of the input through the cipher network, detecting areas of non-random behavior [2]. Such properties are exploited to recover the secret cryptographic key of the cipher. Biham and Shamir presented a novel differential cryptanalysis method [14] that could be applied to various DES-like substitution and permutation cryptosystems, such as FEAL-4 [15]. A classical differential attack follows the exhaustive approach of creating a difference distribution table. Aron Gohr proposed a novel neural network-based distinguisher in his recent work [16], wherein a low-data, chosen-plaintext attack on round reduced SPECK 32/64 gave better results than past work by Dinur on SPECK [17]. Their proposed attack is an all-in-one approach with the Markov assumption which considers all output differences for a given input difference. He also presented a key recovery attack on 11 rounds of SPECK32/64 to recover the last two subkeys after $2^{14.5}$ chosen-plaintext queries with a computational requirement of $2^{38}$ SPECK encryptions, compared to past work by Dinur that achieved the complexity of $2^{46}$ for 11 rounds. Following Gohr's work, Baksi et al proposed a deep learning-based approach for differential attacks on the non-Markov 8-round Gimli-Hash and Gimli-Cipher [18]. They used multiple models including multilayer perceptron (MLP), Long Short-Term Memory (LSTM) [19], and Convolutional Neural Networks (CNN) [20] with varied width and number of neurons. We discuss their method in more detail in a later section.

PRESENT was developed in 2007 [21] at Orange Labs, France. It has become the criteria to measure the security of modern lightweight ciphers [22]. The Simeck family of lightweight block ciphers was developed in 2015 [23] and is a combination of the SIMON and SPECK block cipher families [24], which are the smallest hardware and software block ciphers respectively. It has a smaller hardware footprint and software implementation than its parent families. An overview of the PRESENT and Simeck ciphers is discussed in a later section. These ciphers find practical application in IoT devices such as RFID tags [25] and sensors. Wang [26] in 2008, presented her differential cryptanalysis of 16-round reduced PRESENT. Her proposed differential characteristics for 14 and 15 rounds of encryption had a probability of $2^{-62}$ and $2^{-66}$ respectively. Wang also searched for iterative characteristics from the $2^{nd}$ to $7^{th}$ rounds which she claims to be more effective than the 2-round iterative characteristics. Wang's study proved the effectiveness of using select input differentials which have a higher probability over differentials with lower probability.

This study extends the work by Baksi and Wang for use in two structurally different lightweight block ciphers, PRESENT and Simeck. Our major contributions are as follows:

1. We improve the differential distinguisher algorithm proposed by Baksi et al by using Wang's high probability input differentials for PRESENT and Simeck.
2. We generate three good input differentials from one of Wang's high probability input differences by using left and right shift operators.
3. We propose a simpler deep learning architecture with a lower number of parameters as compared to Baksi's recommended architecture.
4. Our results provide a minimum-security requirement for the discussed ciphers to ensure safety against deep learning-based differential attacks.

Section-2 gives an overview of the lightweight block ciphers PRESENT and Simeck. This is followed by a discussion on the differential distinguisher algorithm in Section-3. We will then discuss our deep learning model followed by the results obtained during experimentation in Sections-4 and 5 respectively. Finally, we conclude the study in Section-6 and provide direction for future work based on our research.

# 2 Lightweight Block Ciphers

Block ciphers are symmetric key ciphers employing deterministic algorithms to encrypt blocks of plaintext. Lightweight block ciphers are a subset of these, as they use algorithms that require less computing power. The following subsections discuss the two lightweight block ciphers used in our study, PRESENT and Simeck.

## 2.1 PRESENT

The PRESENT cipher was developed by Bogdanov et al in 2007 [21]. It was included in the new international standard for lightweight cryptographic methods by the International Organization for Standardization (ISO) and the International Electrotechnical Commission (IEC) in 2019 [22]. Due to its bit-oriented permutations, PRESENT is a hardware cipher which can be implemented with simple wiring. It supports a block-length of 64 bits and key-lengths of 80 bits and 128 bits, which suffice the requirements of moderate-level security applications such as tag-based systems. The scope of this study is limited to the 80-bit key variant of PRESENT.

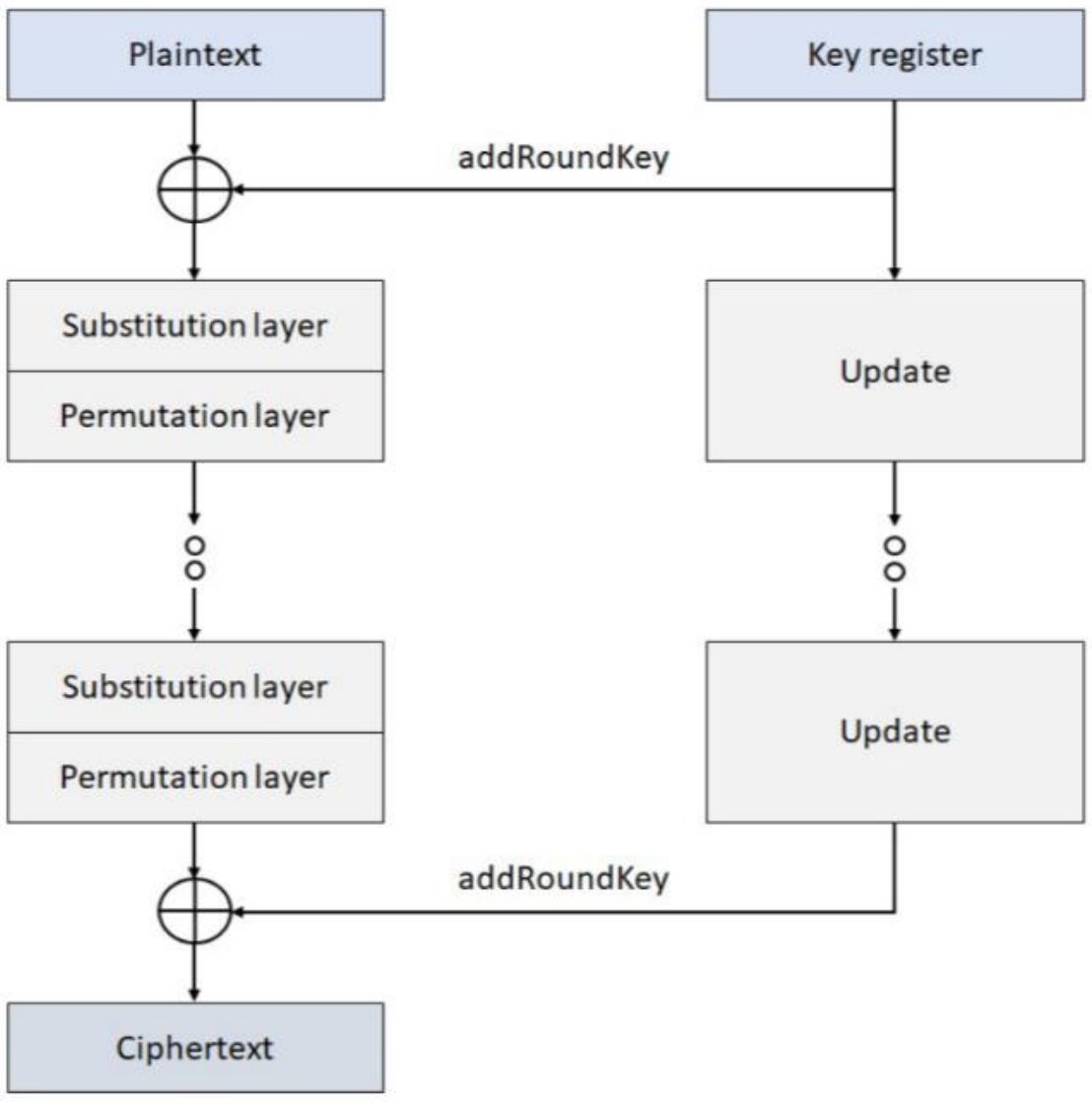

**Fig. 1.** Abstract view of PRESENT cipher.

### 2.1.1 Round Function

A complete-round PRESENT implementation consists of 31 rounds of encryption and each round includes the following layers:

**Substitution Layer:** The substitution layer contains 4×4 S-Boxes. These S-Boxes are used 16 times in parallel each round. The hexadecimal S-Box mapping is shown in Table-1.

**Table 1.** Hexadecimal S-Box Mapping.

| **X** | 0 | 1 | 2 | 3 | 4 | 5 | 6 | 7 | 8 | 9 | A | B | C | D | E | F |
|---|---|---|---|---|---|---|---|---|---|---|---|---|---|---|---|---|
| **S(x)** | C | 5 | 6 | B | 9 | 0 | A | D | 3 | E | F | 8 | 4 | 7 | 1 | 2 |

**Permutation Layer:** Bit-wise permutation is performed on the data block in this layer. The permutation layer mapping is shown in Table-2.

**Table 2.** Permutation Layer Mapping.

| **i** | 0 | 1 | 2 | 3 | 4 | 5 | 6 | 7 | 8 | 9 | 10 | 11 | 12 | 13 | 14 | 15 |
|---|---|---|---|---|---|---|---|---|---|---|---|---|---|---|---|---|
| **P(i)** | 0 | 16 | 32 | 48 | 1 | 17 | 33 | 49 | 2 | 18 | 34 | 50 | 3 | 19 | 35 | 51 |
| **i** | 0 | 1 | 2 | 3 | 4 | 5 | 6 | 7 | 8 | 9 | 10 | 11 | 12 | 13 | 14 | 15 |
| **P(i)** | 0 | 16 | 32 | 48 | 1 | 17 | 33 | 49 | 2 | 18 | 34 | 50 | 3 | 19 | 35 | 51 |
| **i** | 0 | 1 | 2 | 3 | 4 | 5 | 6 | 7 | 8 | 9 | 10 | 11 | 12 | 13 | 14 | 15 |
| **P(i)** | 0 | 16 | 32 | 48 | 1 | 17 | 33 | 49 | 2 | 18 | 34 | 50 | 3 | 19 | 35 | 51 |
| **i** | 0 | 1 | 2 | 3 | 4 | 5 | 6 | 7 | 8 | 9 | 10 | 11 | 12 | 13 | 14 | 15 |
| **P(i)** | 0 | 16 | 32 | 48 | 1 | 17 | 33 | 49 | 2 | 18 | 34 | 50 | 3 | 19 | 35 | 51 |

**addRoundKey:** The round key is XORed to the current block state in accordance with the following equations.

Where $1 \leq i \leq 32$, and $K_i$, $B_i$ are the $i^{th}$ round key and block state respectively. Fig-1 illustrates the general working of the PRESENT cipher.

### 2.1.2 Key Scheduling Algorithm

The round keys to be used in the addRoundKey step are generated using the Key Scheduling Algorithm (KSA) which takes the 80-bit key denoted as $K = k_{79}k_{78}k_{77} \ldots k_0$ and stores it in the shift register. The key is rotated to the left by 61 bits positions. The most significant 4 bits are passed through the 4 × 4 PRESENT S-box. Bits 15-19 of K are then XORed with the least significant bit of roundCounter on the right. The equations for the process are as follows:

$$[k_{79}k_{78} \ldots k_1k_0] = [k_{18}k_{17} \ldots k_{20}k_{19}]$$
$$[k_{79}k_{78}k_{77}k_{76}] = S[k_{79}k_{78}k_{77}k_{76}]$$
$$[k_{19}k_{78}k_{17}k_{16}k_{15}] = [k_{19}k_{18}k_{17}k_{16}k_{15}] \oplus roundCounter$$

### 2.2 Simeck

The Simeck lightweight block cipher was proposed by Yang et al in 2015 [23]. It is designed to be compact and moderately secure for resource constrained applications such as passive RFID tags and Wireless Sensor Network (WSN) nodes. It is a combination of the SIMON and SPECK cipher families which are a set of hardware and software lightweight block ciphers respectively. Simeck is smaller than SIMON due to the reduced size of the Round Function and KSA. It combines the best features of its parent cipher families, that are:

a. A modified and compact version of SIMON's round function.

b. A round function for key scheduling, similar to SPECK.

c. A Linear Feedback Shift Register (LSFR)-based constant for simpler key schedule, in line with SIMON.

The Simeck family of ciphers is represented as Simeck*2n/mn*. Here n is the word size, equal to 16, 24 or 32 based on the family, and *mn* denotes the key size which may be 64, 96 or 128 bits long. Thus, the resulting ciphers are Simeck 32/64, 48/96, and 64/128. These size choices aim to fit various applications such as tag-based systems and are included in the specification of the SIMON and SPECK. The round function and key scheduling algorithm have a Feistel structure as shown in Fig-2 and 3.

#### 2.2.1 Round Function

Fig-2 shows the round function for Simeck. It involves the division of the plaintext into left and right words $l_0$ and $r_0$ respectively, where $l_0$ consists of the most significant $n$ bits, and $r_0$ contains the least significant $n$ bits. The round function processes these two words, followed by the concatenation of $l_T$ and $r_T$, where $T$ is the total number of encryption rounds. The round function is as follows:

$$R_{k_i}(l_i, r_i) = (r_i \oplus f(l_i) \oplus k_i, l_i)$$

Where $l_i$ and $r_i$ are as discussed above and $k_i$ is the $i^{th}$ round key. The function $f(z)$ is as follows:

$$f(z) = (((z) AND (z \lll 5)) \oplus (z \lll 1))$$

In the above equations, $\oplus$, $\lll$ and $AND$ represent the Exclusive OR, left rotation and bitwise AND operations respectively.

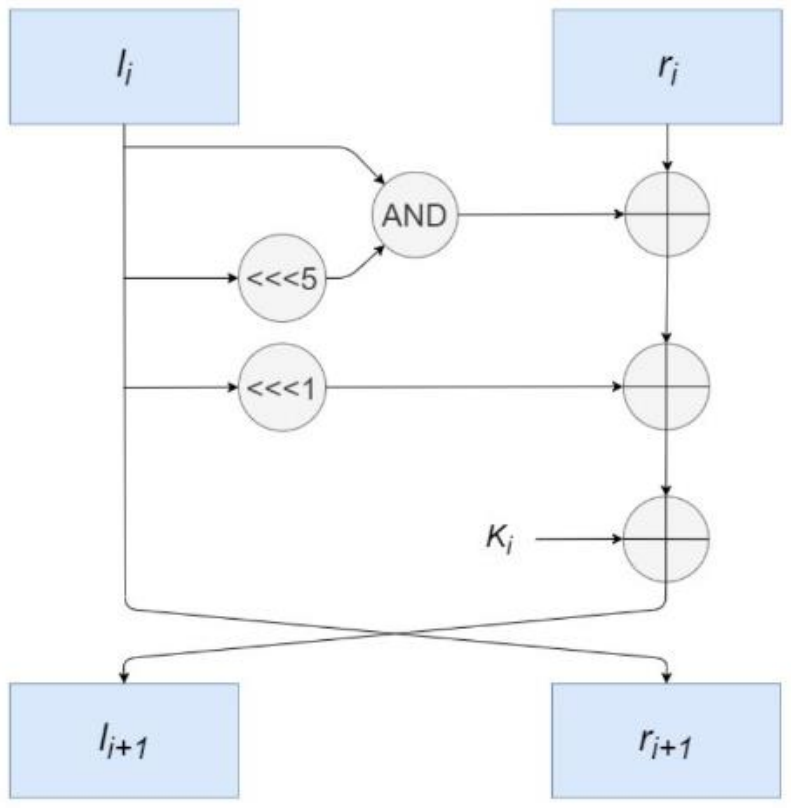

**Fig. 2.** Round function of Simeck.

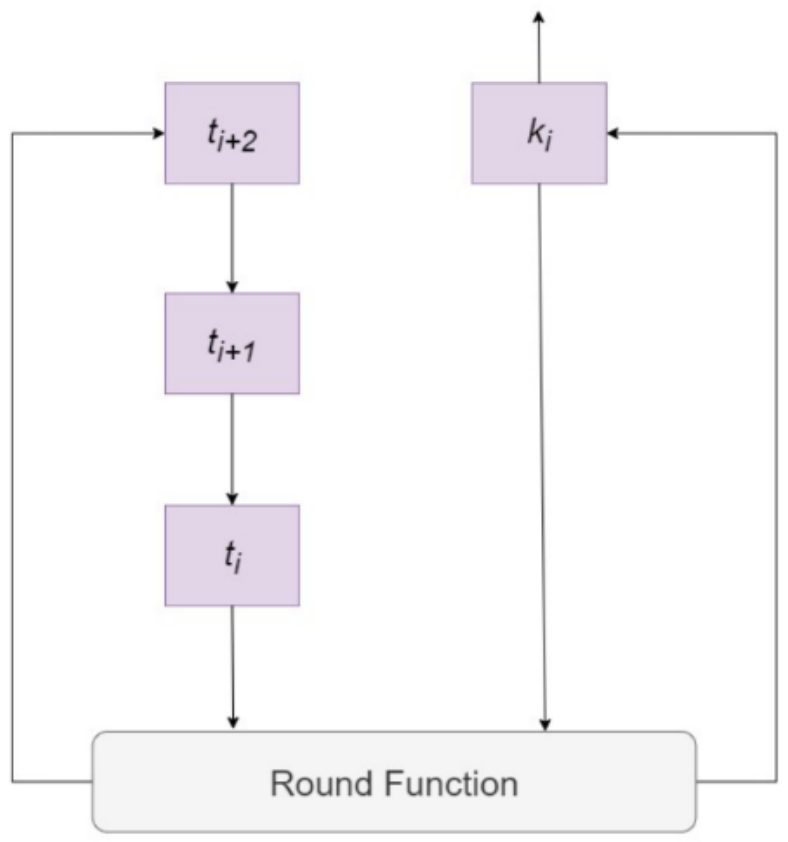

**Fig. 3.** KSA of Simeck.

### 2.2.2 Key Scheduling Algorithm

The $i^{th}$ round key $k_i$ is generated by dividing the master key $K$ into four words and loading them as the inital states $(t_2, t_1, t_0, k_0)$. Here, $t_2$ is the most significant $n$ bits of $k$, and the least significant n bits are denoted by $k_0$. These initial states are loaded into the LFSR. The FSR of the key scheduling algorithm are shown in Fig-3. The update of states is done as per the following equations:

$$k_{i+1} = t_i,$$
$$t_{i+3} = k_i \oplus f(t_i) \oplus C \oplus (z_j)_i$$

Where $i$ is the round number, $C = 2^n - 4$ ($n$ is the word size) and $(z_j)_i$ is the $i^{th}$ bit of sequence $z_j$ in which $j = 0$ for Simeck32/64 and Simeck48/96, and $z_0$ has period of 31. Similarly, for Simeck64/128, $j = 1$ and $z_1$ has a period of 63.

## 3 Differential Distinguisher

Following the work done by Baksi et. al. 18 on Machine Learning (ML) based differential distinguishers, we improve their differential method. This section discusses the in-depth approach of their algorithm and our suggested improvements. The algorithm for the deep learning based differential distinguisher is shown in Algorithm-1. In this differential method, the attacker chooses ($t \geq 2$) input differentials. The four selected differentials are provided in Table-3, where the first differential is selected from Wang's study [26] for PRESENT. The remaining three differentials (numbered 2-4) are generated using left and right shifts. The same differences are used for Simeck64/128 as well. This selection is followed by two phases, OFFLINE and ONLINE. The OFFLINE or the Training Phase is for making the training-dataset of input-output differential pairs, and then training the DL model to learn the relationship between these input and output differences. The ONLINE or Testing Phase involves the creation of the test-dataset and then deciphering whether the given ORACLE is the CIPHER or RANDOM. For $t$ input differentials, if the training accuracy during the Offline phase comes out to be $\geq \frac{1}{t}$, we proceed to the online phase. A testing accuracy $\geq \frac{1}{t}$ in the online phase implies the ORACLE is the CIPHER, and otherwise, RANDOM.

**Table 3.** Differentials used in this study.

| Input differential class | Input differentials |
|---|---|
| 1 | 0x0700000000000700 |
| 2 | 0x7000000000007000 |
| 3 | 0x0070000000000070 |
| 4 | 0x0007000000000007 |

## 4 Deep Learning Model

We have implemented four Differential Distinguisher Models denoted by $M_i$ ($1 \leq i \leq 4$). Model 1 ($M_1$) and Model 2 ($M_2$) use Baksi's recommended deep learning architecture and our proposed architecture respectively, with randomly selected input differentials. Model 3 ($M_3$) and Model 4 ($M_4$) have the same deep learning architectures of $M_1$ and $M_2$ respectively, but with selected differentials. Baksi et al recommend a 4-layer MLP (128, 1024, 1024, $t$). In contrast, our proposed deep learning architecture consists of three layers (128, 1024, $t$), where $t$ is the number of output layer neurons given by the number of differential classes. Fig-4 shows our proposed architecture.

**Algorithm 1** Improved differential distinguisher algorithm.

1: **procedure** TRAINING PHASE (*OFFLINE*)
2: $\delta_t \leftarrow Selected\ differentials$
3: $P, K \leftarrow$ Random
4: $C \leftarrow$ CIPHER (P, K)
5: $\boldsymbol{for}$ $0 \leq i < t$ **do**
6: $C_i \leftarrow$ CIPHER ($P \oplus \delta_i$, K)
7: $dataset(i) \leftarrow$ ($C \oplus C_i, i$)
8: **end** $\boldsymbol{for}$
9: $\boldsymbol{for}$ each $data$ in $dataset$ **do**
10: $\alpha \leftarrow train(data)$
11: **if** $\alpha > \frac{1}{t}$ **then**
12: **goto** *ONLINE PHASE*
13: **else**
14: Repeat from Step 3
15: **end if**
16: **end** $\boldsymbol{for}$
17: **end procedure**

1: **procedure** TESTING PHASE (*ONLINE*)
2: $\delta_t \leftarrow Selected\ differentials$
3: $P, K \leftarrow$ Random
4: $C \leftarrow$ ORACLE (P, K)
5: $\boldsymbol{for}$ $0 \leq i < t$ **do**
6: $C_i \leftarrow$ ORACLE ($P \oplus \delta_i$, K)
7: $dataset(i) \leftarrow$ ($C \oplus C_i, i$)
8: **end** $\boldsymbol{for}$
9: $\boldsymbol{for}$ each $data$ in $dataset$ **do**
10: $\alpha' \leftarrow train(data)$
11: **if** $\alpha' = \alpha$ **then**
12: ORACLE = CIPHER
13: **else**
14: ORACLE = RANDOM
15: **end if**
16: Repeat from Step 3 if required
17: **end** $\boldsymbol{for}$
18: **end procedure**

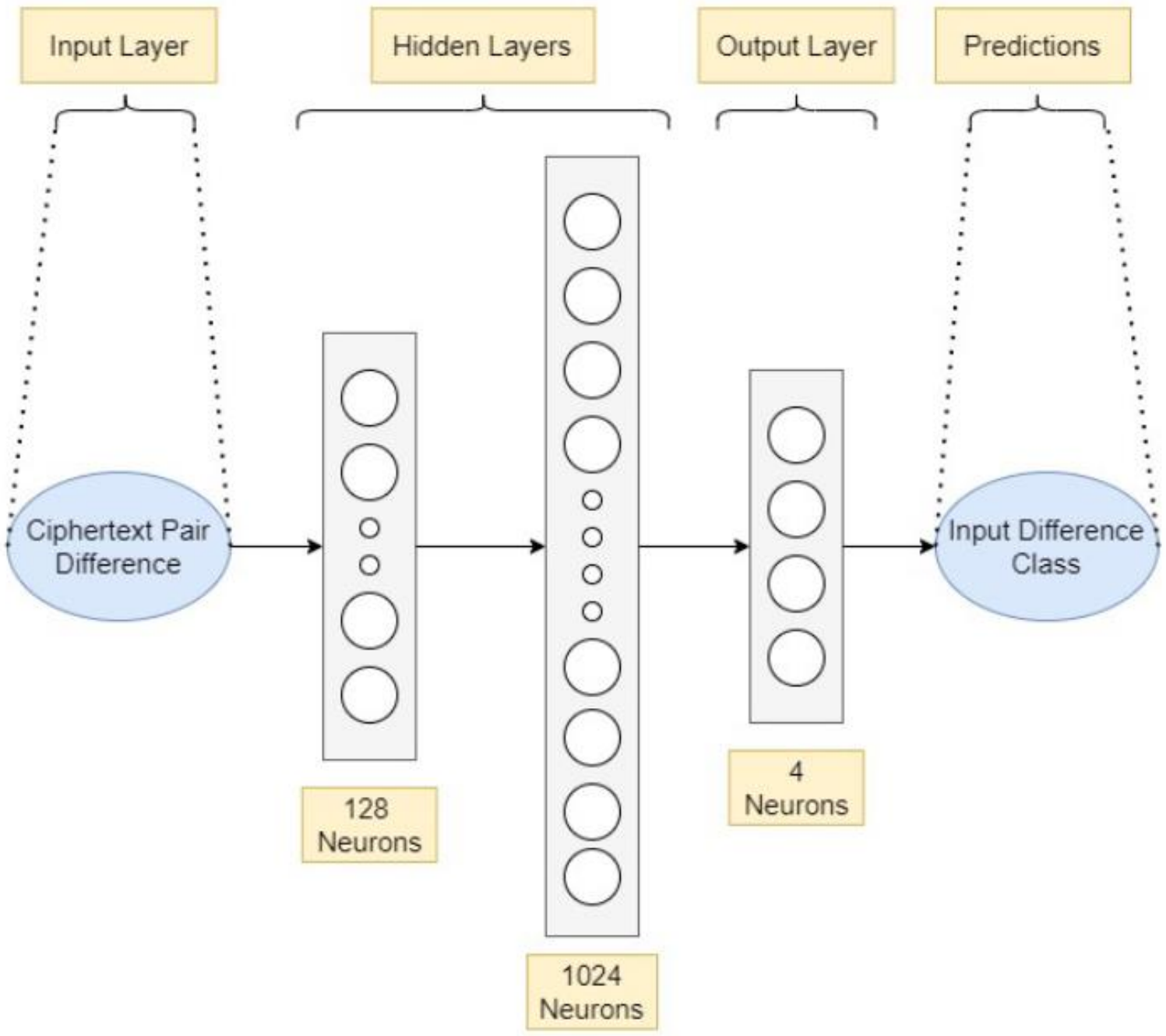

**Fig. 4.** Proposed DL Architecture.

***Dataset Collection***: For 10,000 different key-plaintext pairs, we have selected four input differential classes. These input differentials were either selected randomly (for $M_1$and $M_2$) or were taken from Table-3 (for $M_3$ and $M_4$). For every key-plaintext pair, the plaintext, and its corresponding difference pair, calculated for each input difference class, is encrypted for r-round reduced PRESENT ($3 \leq r \leq 6$) and Simeck ($3 \leq r \leq 7$). The obtained ciphertext pairs XORed to get the output difference. This output difference, along with the input difference class, is stored in a training-dataset. The hyper-parameters for the training and testing of the models are given in Table-4.

**Table 4.** Hyper-parameters.

| **Hyper-Parameters** | **Values** |
|---|---|
| Learning rate | 0.001 |
| Optimizer | Adam |
| Loss function | BCE Loss |
| Epochs | 25 |
| Sample size | 10000 |
| Batch size | 100 |
| Validation Split | 0.3 |

## 5 Results

The previously discussed differential distinguisher models were trained and tested on random data, and data from PRESENT and Simeck. The minimum and maximum validation results of the study are presented in Table-5 in a very concise format.

**Table 5.** Comparison of Differential Distinguisher Models based on Validation Accuracy.

| PRESENT Rounds | Model 1 ($M_1$) | | Model 2 ($M_2$) | | Model 3 ($M_3$) | | Model 4 ($M_4$) | |
|---|---|---|---|---|---|---|---|---|
| | Min | Max | Min | Max | Min | Max | Min | Max |
| 3 | 0.36 | 0.55 | 0.45 | 0.78 | 0.7 | 0.92 | 0.73 | 0.92 |
| 4 | 0.17 | 0.34 | 0.18 | 0.33 | 0.58 | 0.82 | 0.67 | 0.85 |
| 5 | 0.16 | 0.34 | 0.21 | 0.33 | 0.28 | 0.45 | 0.31 | 0.44 |
| 6 | 0.16 | 0.36 | 0.13 | 0.31 | 0.18 | 0.34 | 0.18 | 0.36 |
| **SIMECK Rounds** | **Model 1 ($M_1$)** | | **Model 2 ($M_2$)** | | **Model 3 ($M_3$)** | | **Model 4 ($M_4$)** | |
| | Min | Max | Min | Max | Min | Max | Min | Max |
| 2 | 0.55 | 0.79 | 0.52 | 0.79 | 1.00 | 1.00 | 1.00 | 1.00 |
| 3 | 0.21 | 0.31 | 0.22 | 0.36 | 0.99 | 1.00 | 0.97 | 1.00 |
| 4 | 0.17 | 0.32 | 0.2 | 0.32 | 0.74 | 0.88 | 0.74 | 0.9 |
| 5 | 0.21 | 0.33 | 0.19 | 0.33 | 0.37 | 0.56 | 0.42 | 0.57 |
| 6 | 0.15 | 0.35 | 0.16 | 0.36 | 0.16 | 0.37 | 0.19 | 0.38 |

$M_1$ and $M_2$ did not show any significant validation accuracies as expected due to the random nature of the differentials selected for them. On the other hand, $M_3$ and $M_4$ performed significantly better for both PRESENT and Simeck. This is because the selected input differentials have a high probability for differential cryptanalysis and remove the possibility of using low probability differentials, that would give poor results as seen for $M_1$ and $M_2$. Although the selected differentials are for PRESENT, they give good results for Simeck as well as shown in the table. Fig-5 and Fig-6 show the average validation accuracies for PRESENT and Simeck respectively, round and model wise.

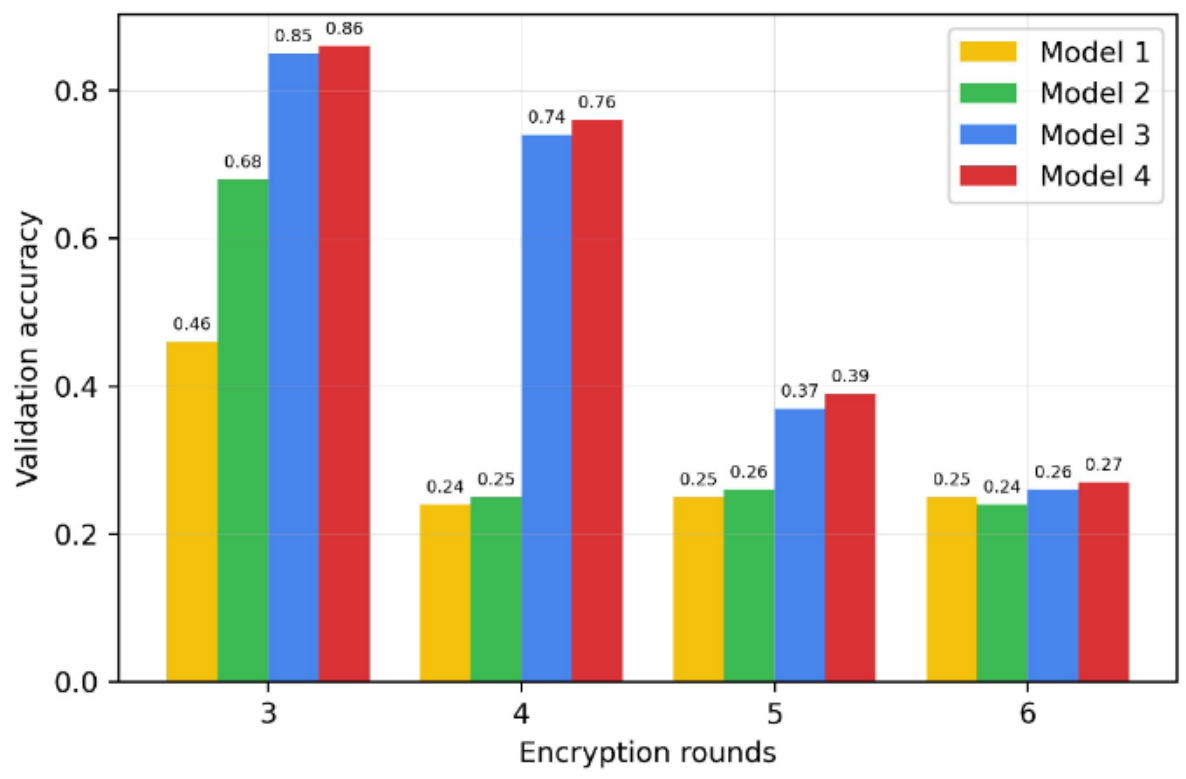

**Fig. 5.** Average validation accuracies for PRESENT.

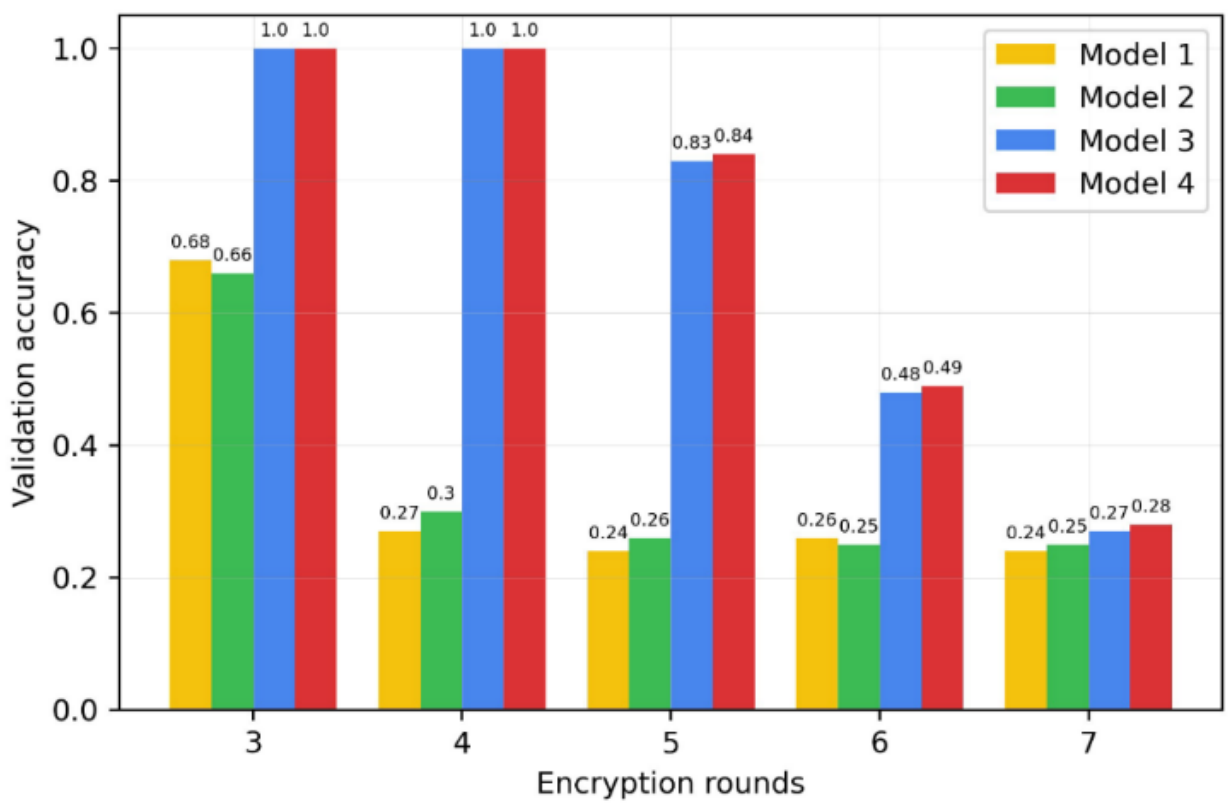

**Fig. 6.** Average validation accuracies for Simeck.

It can be seen that $M_3$ and $M_4$ perform exceptionally well, with obtained average accuracies following a decreasing trend from lower to higher rounds. Whereas the average accuracies for $M_1$ and $M_2$ are significantly lower in comparison, for reasons discussed earlier, with nearly constant values obtained after the third round. All four models reach the limiting average accuracies of nearly 25% by the 6 $6^{th}$ round for PRESENT, and the $7^{th}$ round for Simeck. In addition to the above, our proposed MLP (used in $M_2$ and $M_4$) performs better than Baksi's recommended MLP (used in $M_1$ and $M_3$ ) for use on PRESENT and Simeck.

## 6 Conclusion

In this study, we improved Baksi et al's differential distinguisher method by using one high probability input differential from Wang's study for PRESENT and generated three more differentials from it by shifting it left and right. This improvised method was extended to two structurally different lightweight block ciphers PRESENT and Simeck. The results obtained were significantly better than when randomly selected input differentials were used, and slightly better for our proposed deep learning architecture over the one recommended in the past work. The proposed method can differentiate random data from cipher data until 6 rounds of PRESENT and 7 rounds of Simeck encryption. This shows that using a higher number of encryption rounds for these ciphers can provide the necessary security for IoT devices against a deep learning based differential attack. Based on our study, future work in this area of research can be done on the suggestive universality of the method and input differences across structurally different ciphers. The suggested method for selecting input differentials by shifting an already established high probability input differential can also be explored. In addition to this, our approach does not include a key retrieval method, which can be developed in the future.